\documentstyle[aps,twocolumn,prl,graphicx]{revtex}

\begin{document}

\def\v2o3{V$_2$O$_3$}

\twocolumn[\hsize\textwidth\columnwidth\hsize\csname @twocolumnfalse\endcsname

\title{Orbital occupation, local spin and exchange interactions in \v2o3}

\author{S. Yu. Ezhov$^1$, V. I. Anisimov$^1$, D. I. Khomskii$^2$ and
        G. A. Sawatzky$^2$}

\address{$^1$Institute of Metal Physics, Russian Academy of Sciences,
620219, Ekaterinburg, GSP-170, Russia\\ $^2$Laboratory of Applied and
Solid State Physics, Materials~Science~Centre,
University~of~Groningen, Nijenborgh~4, 9747~AG~Groningen,
The~Netherlands } 
\date{\today} 
\maketitle

\begin{abstract}
\noindent
We present the results of an LDA and LDA+U band structure study of the
monoclinic and the corundum phases of \v2o3 and argue that the most
prominent (spin $\frac{1}{2}$) models used to describe the
semiconductor metal transition are not valid.  Contrary to the
generally accepted assumptions we find that the large on site Coulomb
and exchange interactions result in a total local spin of 1 rather
than $\frac{1}{2}$ and especially an orbital occupation which removes
the orbital degeneracies and the freedom for orbital ordering.  The
calculated exchange interaction parameters lead to a magnetic
structure consistent with experiment again without the need of orbital
ordering. While the low-temperature monoclinic distortion of the
corundum crystal structure produces a very small effect on electronic
structure of \v2o3, the change of magnetic order leads to drastic
differences in band widths and band gaps. The low temperature
monoclinic phase clearly favors the experimentally observed magnetic
structure, but calculations for corundum crystal structure gave two
consistent sets of exchange interaction parameters with nearly
degenerate total energies suggesting a kind of frustration in the
paramagnetic phase. These results strongly suggest that the phase
transitions in \v2o3 which is so often quoted as the example of a
S=$\frac{1}{2}$ Mott Hubbard system have a different origin.  So back
to the drawing board!
\end{abstract}

\pacs{71.27.+a, 71.20.-b, 75.30.Et}
]

\begin{figure} 
 \centering
 \includegraphics[width=7cm]{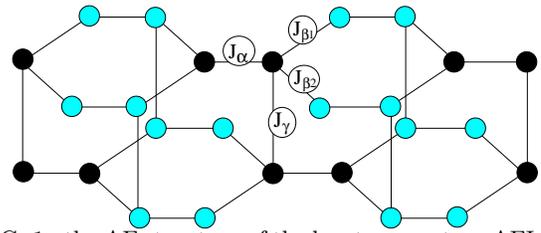} 
 \caption{the AF structure of the low temperature AFI phase of
 \v2o3. The gray and filled circles correspond
 to spin-up and spin-down orientations of the local
 magnetic moments on V ions. The definitions of notations used for the
 superexchange interactions along the various paths are also shown.}
 \label{fig:v2o3_real-afm_struc}
\end{figure}

The \v2o3 system has been a topic of intense study for over more than
fifty years by both theoreticians and experimentalists because of it's
``rich'' phase diagram. It undergoes a first order metal-insulator
transition with a seven orders of the magnitude change in the
electrical conductivity\cite {Foex46}, which can be induced by
temperature, pressure, alloying or by nonstoichiometry. It also
exhibits an antiferromagnetic inslulator to paramagnetic insulator
transition which also is first order and a first order paramagnetic
insulator to paramagnetic metal transition. Many theoretical models
have been put forward to understand the electronic and magnetic
behavior of this compound. Goodenough proposed a model involving both
itinerant and localized $3d$ orbitals \cite{Goodenough70,Goodenough72}
to describe the the \v2o3 electronic structure. Various models have
been suggested and worked out in some detail using mainly the
Mott-Hubbard picture\cite{McWhan69,Zeiger75,Spalek87}. In fact \v2o3
is nowadays used as the best studied example of a Mott Hubbard system
with a semiconductor to metal transition.  To explain the peculiar
antiferromagnetic order (Fig.\ref{fig:v2o3_real-afm_struc}) in the low
temperature insulating phase (AFI) with pairs of parallel spins
coupled antiferromagnetically in the basal plane the intriguing idea
of orbital ordering in the presumably doubly degenerate $E_g$ orbitals
was suggested\cite{Castellani78,Castellani78a,Castellani78b} and
revived recently\cite{Rice95} to explain new neutron scattering
results\cite{AeBa:95}. Also within the context of infinite dimension
calculations to describe strongly correlated systems \v2o3 is now used
as a good example of the success of this
approximation\cite{Kotliar95}.  However, the character of the phase
transition and the nature of the ground state is still very
controversial. Recent photoemission and X-Ray absorption
results\cite{Tjeng99} strongly suggest that the AFI ground state
should not be described as a spin $\frac{1}{2}$ antiferromagnetic
Mott-Hubbard insulator as assumed in the above theories and, related
to this, that the $d$ orbital occupation is quite different from that
conventionally assumed.  In this letter we present the study of the
electronic structure of \v2o3 in LDA\cite{HoKo:64,Kohn65} and
LDA+U\cite{Anisimov91,Anisimov97} approximations. In contrast to the
standard LDA, in LDA+U method the influence of the on-site $d$-$d$
Coulomb interaction is included in Hamiltonian, as an effective
potential which is different for electron removal than for electron
addition which has been shown to be crucial to describe strongly
correlated materials.  

\begin{figure}
 \includegraphics[angle=270,width=8cm]{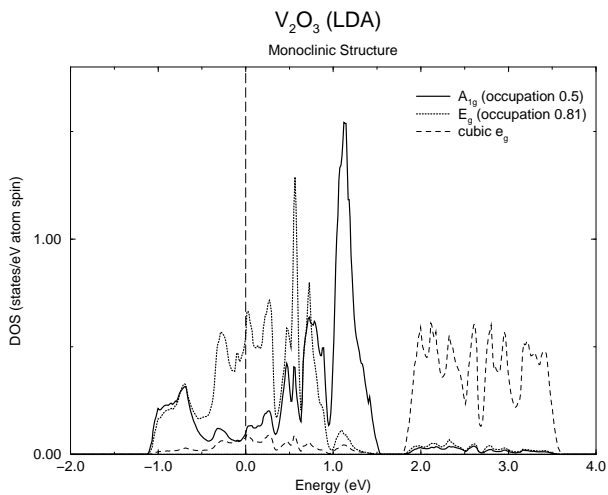}
 \caption{Partial densities of state for $3d$ states of V obtained in
 the LDA calculation.  Occupation numbers are given per one orbital
 (and both spins), i.e. each of the $E_g$ orbitals has occupation of
 0.81 electrons. The Fermi level is at zero
 energy.}
 \label{fig:v2o3_lda_m}
\end{figure}

The crystal structure in the low temperature AFI phase is monoclinic
\cite{Dernier70a}, above $T_c$ this changes to the corundum
structure. (T$_c\simeq 150K$\cite{Dernier70}). The calculated
densities of states for both structures are very similar showing that
the electronic structure itself is little influenced by the lattice
distortions so that the strong change in properties must be a rather
subtle effect with respect to the electronic structure. In
Fig.\ref{fig:v2o3_lda_m} the partial densities of states (DOS) obtained
in the LDA calculation for the monoclinic crystal structure of \v2o3
are shown. They are very similar to those found by Mattheis's
\cite{Mattheis94} for the corundum lattice. As the monoclinic
distortion of the lattice is not very strong, we can plot DOS's
assuming approximate trigonal symmetry. The V ions are somewhat off
center in a slightly trigonally distorted octahedron of O ions. This
distortion causes the otherwise 3 fold degenerate $t_{2g}$ 3d orbitals
to split into a non-degenerate $A_{1g}$ and double degenerate $E_g$
levels. In this representation the $A_{1g}$ orbital has $3z^2-r^2$
symmetry in a hexagonal coordinate system, i.e. with z axis along $c$
direction (V-V pairs), and $E_g$ orbitals are directed more towards
the V ions in the basal plane.

At first glance the LDA picture looks very simple and one may conclude
that the conventionally used ideas concerning the splitting of the
$A_{1g}$ orbitals into bonding and antibonding partners because of a
hopping integral between the V ions in the pairs along the $c$ axis
with two electrons per pair in the bonding orbital are
confirmed. However upon closer examination a different picture emerges
which is further supported by the LDA+U calculations below as well as
the recent experimental XAS results\cite{Tjeng99}. First we see indeed
a rather broad $A_{1g}$ band with a total width of a little more than
2~eV and an about $\frac{1}{2}$~eV narrower $E_g$ band both
stradelling the Fermi energy and resulting in a metallic state. The
$A_{1g}$ band, although it shows some structure that might be
interpreted as a bonding-antibonding pseudo gap, actually exhibits a
strong peaking above $E_f$ and only relatively little weight below
$E_f$. This looks very different from a bonding-antibonding splitting
which should have exhibited a more symmetric structure about the
center of gravity if dimer hoping integrals were dominating the
problem. We also note that the total $A_{1g}$ band width is only
2.5~eV which is considerably smaller than the expected value of the
Hubbard U of about 3~eV including the screening due to the strongly
bonding $e_g$ electrons\cite{Solovyev96} or about 4-5 eV without
this screening channel\cite{Mizokawa93}. Such a small band width
would invalidate a molecular orbital like approach also for the
$A_{1g}$ orbitals. So even these results already cast doubt on the
validity of the most commonly used starting point with a strong
bonding-antibonding splitting of the $A_{1g}$ orbitals, with the $E_g$
orbitals in this gap ending up with two spin-antiparallel electrons
per $c$-axis pair in the $A_{1g}$ bonding state leaving only one
electron in an assumed narrow doubly degenerate $E_g$
band\cite{Castellani78,Castellani78a,Castellani78b}. It was this
starting point which led to the now very much used one electron per
site Hubbard model for \v2o3.

The total occupation number of the $A_{1g}$ band is only 0.5 electrons
per V atom. Another very important point, which can be derived from
the LDA result is that the $E_g$-$A_{1g}$ actual splitting due to the
trigonal crystal field is as large as $\sim$0.4 eV with the $E_g$ band
center of gravity lower than the $A_{1g}$ band center of gravity. This
large splitting makes the $E_gE_g$ configuration for the localized AFI
phase more favorable than the $A_{1g}E_g$ configuration removing the
orbital degeneracy. We can clearly see at this point that if we switch
on the on-site Coulomb interaction $U$ in our LDA+U calculations the
$A_{1g}$ states will be pushed up above the Fermi level and the
resulting ground state will be $E_gE_g$. This implies that the ground
state (AFI phase) will not be degenerate leaving no place for any
orbital ordering in the usual sense.

\begin{figure}
 \centering \includegraphics[angle=270,width=8cm]{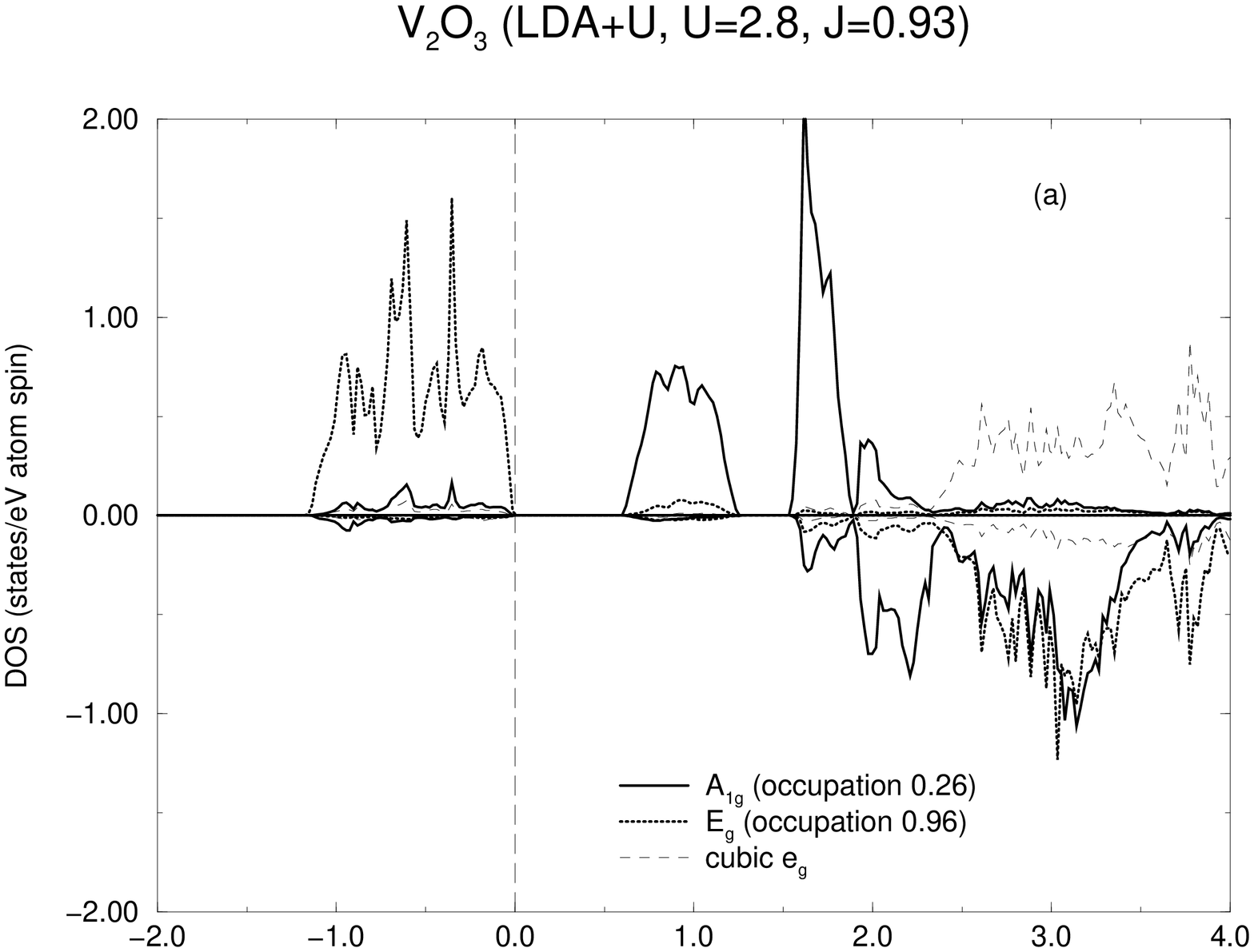}
 \includegraphics[angle=270,width=8cm]{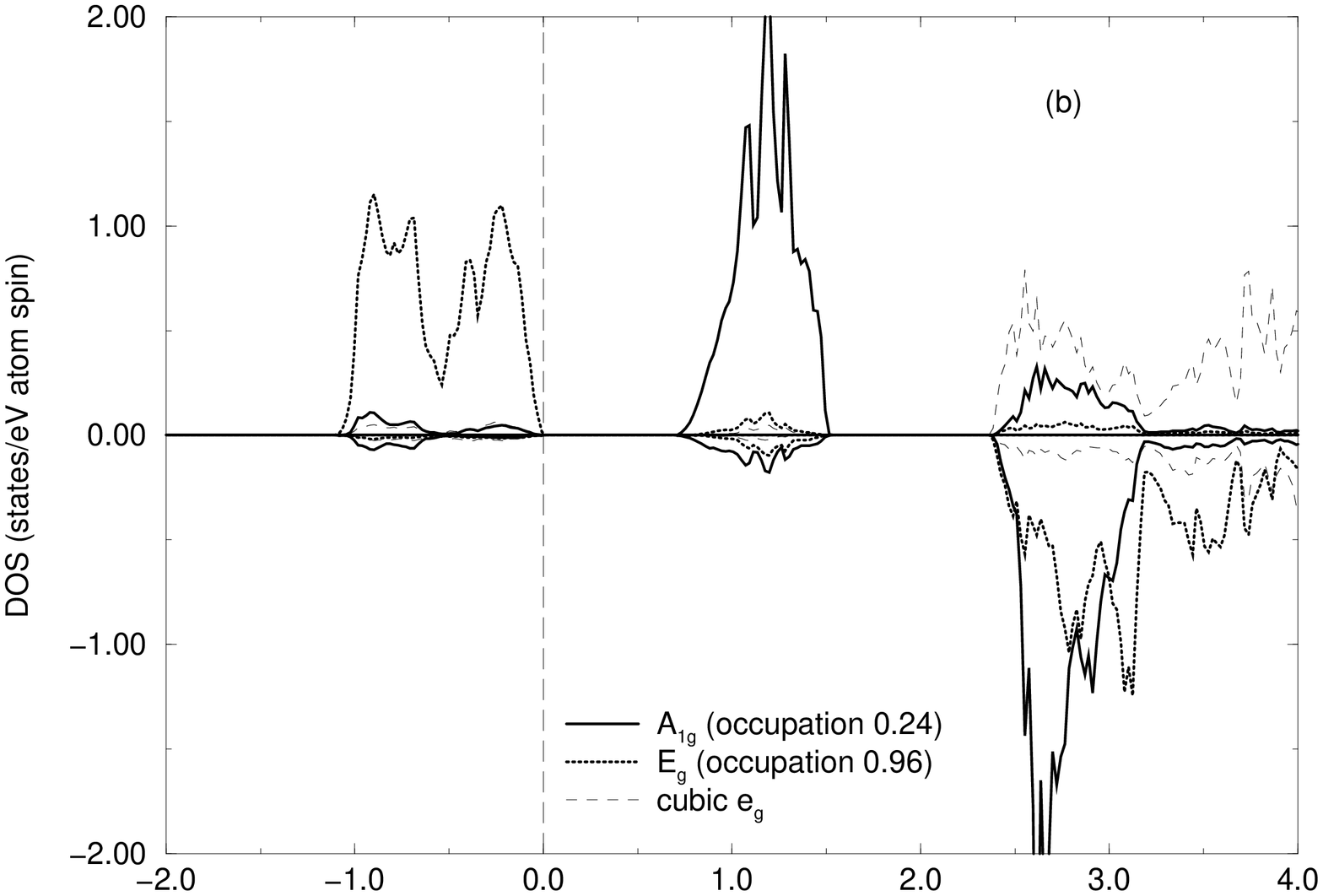}
 \includegraphics[angle=270,width=8cm]{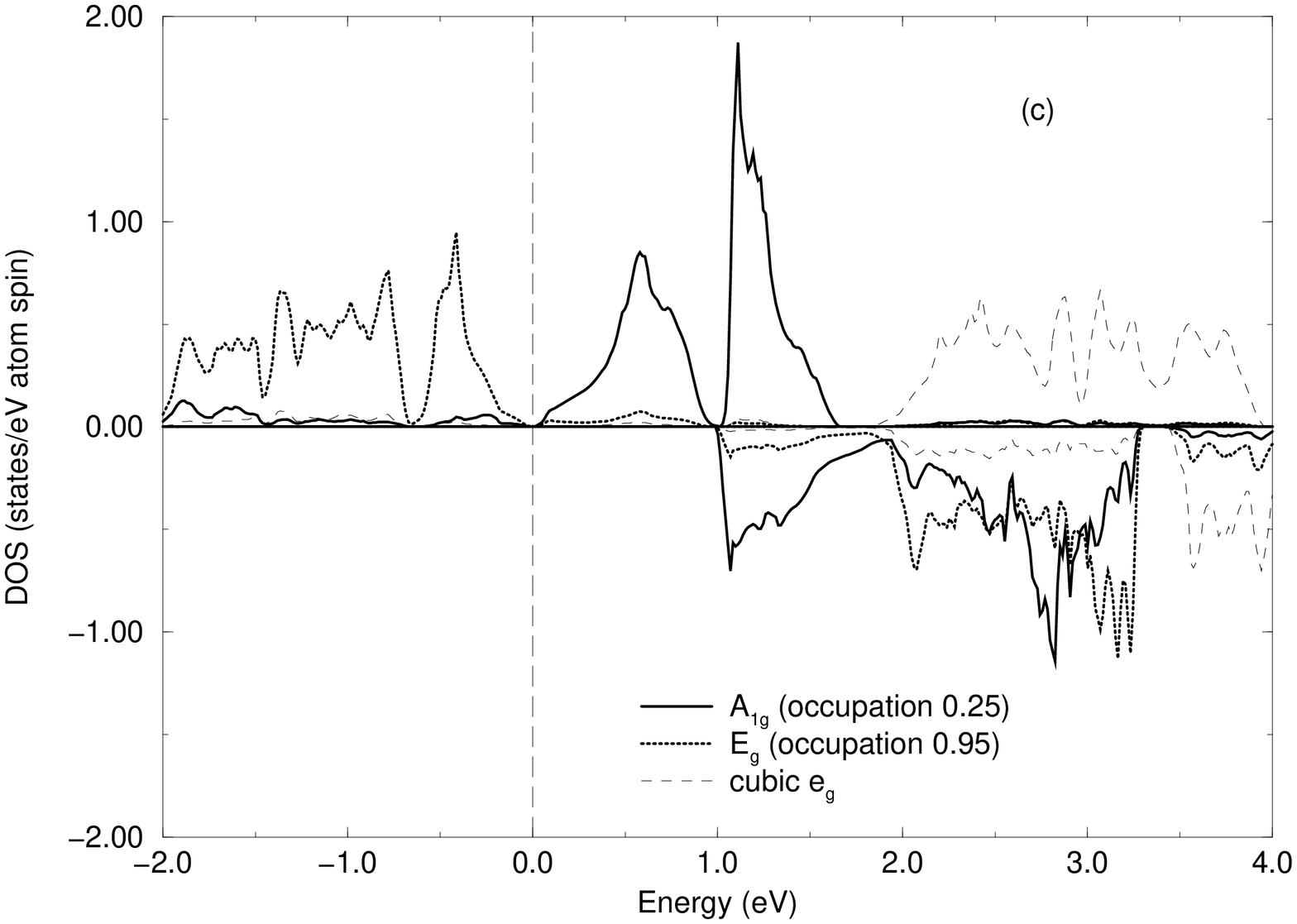} 
 \caption{Partial densities of states for $d$ states of V obtained in
          LDA+U calculation. a) ``Real'' antiferromagnetic structure; b)
          ``simple'' antiferromagnetic structure; c:) ferromagnetic structure}
 \label{fig:v2o3_lda+u_m}
\end{figure}

The above LDA results give us rough ideas which should be further
checked. The LDA+U calculations can provide us with some more details
about the electronic structure of \v2o3.  We performed LDA+U
calculations for different magnetic structures, namely the ``real'' AF
(Fig.\ref{fig:v2o3_real-afm_struc}), a ``simple'' AF (all the nearest
neighbors are antiferromagnetically coupled), a ``layered'' AF (all
the nearest neighbors in basal plane are antiferromagnetically coupled
and the neighbor along hexagonal $c$ axis is coupled
ferromagnetically) and FM structure.  In Fig.~\ref{fig:v2o3_lda+u_m}
the LDA+U partial DOS are plotted. Here the results for U=2.8 (eV) and
J=0.93 (eV) are presented. These values of U and J were calculated by
Solovyev {\em et.~al.}, taking into account the screening of $t_{2g}$
interactions by $e_g$ electrons\cite{Solovyev96}. The striking point
of the LDA+U result is that the electronic structure strongly depends
on the magnetic structure. The band gap for instance in the real AF
magnetic structure is 0.6 (eV) close to the experimental value
[Fig.\ref{fig:v2o3_lda+u_m}(a)], while in the FM structure we find a
half-metal[Fig.\ref{fig:v2o3_lda+u_m}(c)]. The nature of the DOS
changes strongly even far below and and above Fermi level indicating
the very strong sensitivity of the electronic structure to the spin
structure. However, the occupation numbers for all magnetic structures
are $\sim 0.95$ for each of $E_g$ orbital, and $\sim 0.25$ for
$A_{1g}$, which formally corresponds to $E_gE_g$ configuration.
In the present calculation
there is no sign of any orbital ordering at all. We have calculated
quadrupole moment tensor for d-shell of V ion and have found that
while 3-fold symmetry is broken for ``real'' AFM structure, the
quadrupole moment tensor is the same on all V ions. 
We should mention that we did not include the spin orbit coupling 
in these calculations and these could have appreciable effects for V$^{3+}$. 
For example the crystal structure allows for a 
Dzyaloshinsky-Moria coupling 
which could result in a small non colinear spin allignement. This 
could in the end increase the unit cell and could be visible in an 
annomalous scattering experiment. 

 We should stress here that the total energy
difference between different magnetic structures is really small. For
the monoclinic crystal structure the ``real'' AF state has the lowest
total energy followed by the simple AF state, which is higher by
80~K. In the case of the corundum crystal structure the simple AF
state has the lowest energy, but the energy difference with the
``real'' AF state was only 5~K.  Thus in the corundum phase these two
magnetic structures (real and simple) turned out to be almost
degenerate. In short, the LDA+U results show us how important the
small monoclinic distortions are for the magnetic structure and how
low energy scale excitations, involving only spin reorientation, can
effectively change large energy scale values such as band gaps.

There are two problems arising from the present results. First the
spin should be considered to be 1 per V atom rather than 1/2 and
secondly that the orbital occupation is consistent with a $E_gE_g$
configuration of the ground state, which is in-plane symmetric and
orbitally nondegenerate, but still with the complex "Real" magnetic
structure of the AFI phase. In this magnetic structure shown in Fig 1,
every atom has three closest neighbors in the basal plane, one of
which is ferromagnetically aligned ($J_\alpha $) and the other two ---
antiferromagnetically ($J_{\beta _1}$ and $J_{\beta _2}$), and also
there is one neighbor along the hexagonal $c$ axis which is
ferromagnetically aligned ($J_\gamma $)(see
Fig.\ref{fig:v2o3_real-afm_struc} for definitions of exchange
interaction parameters). To check the above mentioned consistency we
calculated the exchange interaction parameters (EIP) using a well
tested method described in\cite{Licht95}. We calculated EIP's for both
crystal structures: monoclinic and corundum. It was found that
according to EIP calculation only one stable magnetic structure exists
in the monoclinic phase.  ("Stable" (or "consistent") means that if in
the EIP calculations the certain pair of spin was parallel, the
corresponding exchange parameter came out as ferromagnetic, and if
antiparallel, as antiferromagnetic.) It is the ``real'' AF magnetic
structure with the following values of EIP's: $J_\alpha =48$~K
(ferromagnetic), $J_{\beta _1}=-214$~K, $J_{\beta _2}=-90$~K,
$J_\gamma =47$~K. For the corundum crystal structure we obtained two
stable magnetic configurations from the EIP calculation point of view:
``real'' AF ($J_\alpha =55$, $J_{\beta _1}=J_{\beta _2}=-120$,
$J_\gamma =44$) and magnetic structure with uniform AF exchange in the
basal plane ( $J_\alpha =J_{\beta _1}=J_{\beta _2}=-65$) and small
frustrated exchange along the hexagonal $c$ axis
($J_\gamma\simeq0$). From these results we can say, that indeed the
monoclinic distortion of the crystal structure stabilizes the real AF
magnetic structure, and the $E_gE_g$ configuration of the $d$
electrons is consistent with this magnetic structure. The values of
EIP's depend strongly on the magnetic structure, which tells us that
one cannot adequately model the magnetic interactions in \v2o3 as only
nearest neighbor Heisenberg exchange.  This is most probably connected
with the fact that \v2o3 is close to being metallic which is also
reflected in the strong dependence of the electronic structure on the
magnetic one (and vice versa) which we saw in LDA+U calculation. The
polarized neutron scattering experiments \cite{Bao97} show the
qualitative change of magnetic interactions in the transition from the
antiferromagnetic insulator with the monoclinic crystal structure to
both the metallic phase and paramagnetic insulator with the corundum
structure. Instead of a peak in reciprocal space corresponding to AFI
magnetic structure (``real'' AF) it has a peak corresponding to a
magnetic structure with all three V-V interactions in basal plane
antiferromagnetic (``layered'' AF). Another peculiarity of the neutron
scattering results is a very large width of the peak.

Our results show that for the distorted monoclinic crystal structure
the ``real'' AF magnetic structure is the lowest one and the only one
which gives a consistent set of exchange interaction parameters.
Transition to the corundum crystal structure leads to the coexistence
of the two well defined sets of exchange interaction parameters with
nearly degenerate values of total energy for those two magnetic
configurations, the symmetric in basal plane ``layered'' AF structure
being slightly lower in energy. That would result, in reciprocal
space, in a peak centered at the corresponding Q-vector but this peak
will be strongly broadened due to the transitions to the excited state
and corresponding fluctuations in the magnetic structure.

Our calculations also give a value of the magnetic moment per V
$\sim1.7\,\mu_B$, nearly the same in all the structures studied.  This
value is somewhat larger than the value of $1.2\,\mu_B$ obtained from
neutron scattering for the antiferromagnetic phase, but is consistent
with the value (1.7 $\mu_B$) obtained from the high-temperature
susceptibility.  This relatively large value of $\mu$ is evidently a
consequence of a strong Coulomb interaction on V$^{3+}$ which tends to
destroy the formation of the molecular orbital singlet state on
$A_{1g}$-orbitals of V-V pair, which was assumed in most previous
studies.  The fact that this is independent of the magnetic or crystal
structure suggests that this should be treated as a high energy scale
parameter in any model.

Summarizing, we have carried out LDA+U calculations of the electronic
structure and exchange constants of \v2o3 in both the monoclinic and
corundum structures and obtained a consistent description of the main
properties of the antiferromagnetic insulating phase and of the
paramagnetic insulating one. In contrast to the previous assumptions,
in both these phases the electronic configuration is predominantly
$E_gE_g$ one, i.e. two $d$-electrons of $V^{3+}$ occupy the doubly
degenerate $E_g$-orbitals. In addition the spins of the two electrons
are parallel leading to a high spin S=1 local moment.  As a result
there is no orbital degeneracy left and correspondingly no orbital
ordering of the kind which was invoked previously to explain magnetic
properties of
\v2o3\cite{Castellani78,Castellani78a,Castellani78b,Rice95}.
 Despite that we are able to obtain the correct
magnetic structure of \v2o3: the signs of the exchange constants in
the monoclinic phase are consistent with the observed
antiferromagnetic structure.  The calculated values of the energy gap
$\sim0.6\,$eV and the magnitude of the magnetic moment per V
$\sim1.7\mu_B$ also agree with the experiment.  Strong change of
magnetic correlations through $T_N$ is ascribed to the near degeneracy
of different magnetic structures in the corundum magnetic structure of
\v2o3, the state with antiferromagnetic correlations in all three
directions in a basal plane having slightly lower energy.  Thus
orbital ordering is not required to explain the physical properties of
\v2o3 in the antiferromagnetic and in the paramagnetic insulating
phases. The LDA+U results also demonstrate that a spin 1/2 Hubbard
model is not the correct starting point but that one should use a spin
1 model with very strong hunds rule exchange. Within such a model with
two electrons S=1 in $E_g$ orbitals on each site the hopping would
have to involve both the minority spin $E_g$ orbitals as well as the
$A_{1g}$ orbitals which are close in energy. Having found this it is
also not so surprising any more that the electronic structure and
especially the d band widths and splittings are strongly dependent on
the nearest neighbor spin spin correlation functions and that there
may be a redistribution of electrons between the $E_g$ and $A_{1g}$
states on going into the PI or Metallic phase as found in recent
electron spectroscopic studies\cite{Tjeng99}.

This investigation was supported by the Russian Foundation for Fundamental
Research (RFFI Grant No. 98-02-17275), by the Netherland Organization
for Fundamental Research on Matter (FOM) with financial support by the
Netherlands Organization for the Advance of Pure Science (NWO).

\bibliographystyle{prsty}
\bibliography{journals,v2o3,dft,ezh}

\end{document}